\documentclass[fleqn,usenatbib]{mnras}

\usepackage{newtxtext,newtxmath}

\usepackage{multirow}
\usepackage[table,xcdraw]{xcolor}
\usepackage{arydshln}

\usepackage[T1]{fontenc}

\DeclareRobustCommand{\VAN}[3]{#2}
\let\VANthebibliography\thebibliography
\def\thebibliography{\DeclareRobustCommand{\VAN}[3]{##3}\VANthebibliography}

\usepackage{graphicx}	
\usepackage{amsmath}	

\title[AGN Fueling versus Environment]{The co-evolution of strong AGN and central galaxies in different environments}

\author[Sampaio et al. 2023]{V. M. Sampaio,$^{1,2}$\thanks{E-mail: vitor.sampaio@nottingham.ac.uk}
A. Arag\'on-Salamanca,$^{1}$ M. R. Merrifield$^{1}$, R. R. de Carvalho,$^{2}$ S. Zhou,$^{1}$ I. Ferreras$^{3,4}$   
\\
$^{1}$ School of Physics and Astronomy, University of Nottingham, University Park, Nottingham NG7 2RD, UK\\
$^{2}$ NAT - Universidade Cidade de S\~ao Paulo, 01506-000, SP, Brazil\\
$^{3}$ Instituto de Astrof\'isica de Canarias, Calle V\'ia L\'actea s/n,
E38205, La Laguna, Tenerife, Spain\\
$^{4}$ Department of Physics and Astronomy, University College London, Gower Street, London WC1E 6BT, UK
}

\date{Accepted 2023 July 18. Received 2023 July 11; in original form 2023 March 11}

\pubyear{2023}

\begin{document}
\label{firstpage}
\pagerange{\pageref{firstpage}--\pageref{lastpage}}
\maketitle

\begin{abstract}
We exploit a sample of over 80,000 SDSS galaxies at the centre of their local halo to investigate the effect of AGN feedback on their evolution. We trace the demographics of optically-selected AGN (Seyferts) as a function of their internal properties (stellar and halo mass, central velocity dispersion, star-formation activity, morphology) and environment. We find that the preeminence of AGN as the dominant ionising mechanism increases with stellar mass, overtaking star-formation for galaxies with $M_\text{stellar} \geq 10^{11}M_\odot$. The AGN fraction changes systematically with the galaxies' star-formation activity. Within the blue cloud, this fraction increases as star-formation activity declines, reaching a maximum near the green valley ($\sim 17 \pm 4\%$), followed by a subsequent decrease as the galaxies transition into the red sequence. This systematic trend provides evidence that AGN feedback plays a key role in regulating and suppressing star formation. In general, Seyfert central galaxies achieve an early-type morphology while they still host some residual star formation. This suggests that, in all environments, the morphology of Seyfert galaxies evolves from late- to early-type before their star formation is fully quenched. Stellar mass plays an important role in this morphological transformation: while low mass systems tend to emerge from the green valley with an elliptical morphology (T-Type $\sim -2.5 \pm 0.7$), their high-mass counterparts maintain a spiral morphology deeper into the red sequence. In high-stellar-mass centrals, the fraction of Seyferts increases from early- to late-type galaxies, indicating that AGN feedback is linked with the morphology of these galaxies and, perhaps, its transformation. Our analysis further suggests that AGN are fuelled by their own host halo gas reservoir, but AGN at the centres of groups can also increase their gas reservoir via interactions with satellite galaxies. 
\end{abstract}



\section{Introduction}

The last few decades, astronomers have focused on understanding the different mechanisms driving galaxy evolution across cosmic time. The persistent bimodality observed in the distribution of several galaxy properties such as their colours and star-formation rates \citep{Strateva1, 2006MNRAS.373..469B, 2012MNRAS.424..232W} provides not only a key piece of evidence, but also a major challenge for any theory of galaxy formation and evolution that that tries to explain it. 

Galaxies divide between the blue cloud (BC) -- blue star-forming galaxies dominated by late-type morphologies (late-type galaxies, LTGs); the red sequence (RS) -- red galaxies with little or no star formation, populated mainly by galaxies with early-type morphologies (early-type galaxies, ETGs); and the green valley (GV) -- an intermediate class comprising galaxies with intermediate colours, reduced star formation, and a range of morphologies \citep{2009MNRAS.396..818S, 2010MNRAS.404..792M, 2010MNRAS.405..783M, 2014MNRAS.440..889S, 2019MNRAS.488L..99A}. There have been many attempts at explaining the transition from the BC through the GV to the RS invoking a variety of physical mechanisms to reduce and quench the galaxies' star formation and change their morphology \citep[e.g.][]{2020MNRAS.491.5406T, 2022MNRAS.509.3889D}, but a definitive answer is still to be found. 

A particularly interesting class of objects are the so-called `central galaxies'. These are systems located at that the centre of dark matter potential wells, and therefore at the focus of any possible gas accretion. Because they are the dominant galaxy in their dark-matter halo, it is reasonable to expect that the influence of external objects may be smaller than for satellite galaxies. If this is the case, internal processes may be relatively more dominant than external (environmental) ones, and therefore easier to study.    

A great deal is already known about central galaxies, providing a solid framework to study their transformation. Their evolution has been show to depend on the mass of their host dark-matter halo ($M_\text{halo}$), the velocity dispersion of the cluster/group they inhabit ($\sigma_\text{cluster}$) and the number of its members ($N_\text{members}$) \citep{2003ApJ...585..687K, 2018ApJ...860..102W, 2021A&A...650A.155Z, 2022ApJ...933L..12G}. All these properties correlate with the depth of the potential well at whose centre these galaxies live, linked with their ability to accrete gas. The velocity dispersion and the number of members also relate to the nature of their potential interactions with other galaxies and the probability of merging with them \citep{1943ApJ....97..255C, 1976MNRAS.174...19W, 1977MNRAS.181..735B}.  Furthermore, \cite{2013ApJ...762L..31B} have shown that, in general, star-formation efficiency peaks at $M_\text{halo}\sim 10^{11.7}\text{M}_{\odot}$. However, halo mass is not the only driver of star formation in central galaxies; The detailed properties of their environment also matter. For instance,  \cite{2014MNRAS.445.1977L} find that central early-type galaxies hosted by halos with $N_\text{members} \geq 3$ and $M_\text{halo} \geq10^{12.5}\text{M}_{\odot}$ have younger ages and longer star-formation timescales (lower alpha enhancement, $\rm [\alpha/Fe]$) than isolated centrals occupying less massive halos ($\leq 10^{12.5}\text{M}_{\odot}$). Younger ages and longer star-formation timescales may be the result of gas-rich interactions with satellite galaxies \citep{1992ApJ...387..152J}, whose probability depends crucially on environment. 
 
One of the main \textit{internal} drivers of galaxy evolution is the presence of a central super-massive black hole (SMBH). According to \cite{2013MNRAS.433.3297D} and \cite{2022PhR...973....1D}, the SMBH mass is one of the parameters that most effectively predicts whether a galaxy is star-forming or quenched. It has been suggested \citep[e.g.,][]{2014MNRAS.441.3055B} that the relation between SMBH mass and star-formation suppression is due to the effect an active galactic nucleus (AGN) has on the interstellar medium of the galaxy (often called `AGN feedback'). The effect an AGN has will depend both on the mass of the SMBH at its centre and the amount and rate of `fuel' feeding it. 

Empirically, \cite{1998AJ....115.2285M}, \cite{2000ApJ...539L...9F}, \cite{2000ApJ...539L..13G}, \cite{2002ApJ...574..740T} and \cite{2009ApJ...698..198G} have shown a positive correlation between the SMBH mass and the central stellar velocity dispersion of its host galaxy ($\sigma_\text{galaxy}$), while \cite{2016MNRAS.462.2559B} and \cite{2021A&A...650A.155Z} find an increasing fraction of quenched galaxies with increasing $\sigma_\text{galaxy}$. Since cluster centrals are generally the galaxies with the largest $\sigma_\text{galaxy}$ \citep{2003ApJ...594..225S}, we expect them to host the most massive SMBHs, and therefore show signs of strong star-formation suppression. 

A variety of models try to address the physics behind AGN feedback and its effect on a galaxy's interstellar medium and thus its star formation \citep{1998A&A...331L...1S, 1999MNRAS.308L..39F, 2003ApJ...596L..27K, 2010MNRAS.402.1516K}. High velocity AGN-driven winds and radio jets have been shown to prevent the cooling of the available gas and even to cause gas outflows \citep{2017FrASS...4...42M, 2018MNRAS.480.3993B, 2020A&A...644A..54S}, therefore preventing this gas from forming stars. \cite{2012ApJ...758...66W} also show that radiation pressure around the AGN region can cause a vertical circulation of gas, a phenomenon called a `galactic fountain', which constantly prevents the collapse of gas clouds. These are examples of `negative' AGN feedback, where star formation is suppressed. On the other hand, \cite{1989ApJ...345L..21B} and  \cite{1989MNRAS.239P...1R} propose that the alignment of radio and optical structures in high redshift galaxies is the result of localised enhanced star formation due to the AGN radio jets themselves. This would be a form of `positive' AGN feedback, which enhances star formation instead of suppressing it. \cite{1997ApJ...491...78C} further suggests that the most powerful jets are responsible for the formation of the most massive elliptical galaxies, whereas less powerful ones are linked to smaller ellipsoids and bulges. \cite{2012MNRAS.425..438G} show that positive AGN feedback is also found in simulations. The balance between positive and negative feedback is not fully understood, and the possibility of a combination of both modes has also been debated in the literature \citep{ 2013ApJ...772..112S, 2013ApJ...774...66Z, 2013MNRAS.431..793Z}. In this paper we will try to use a relatively large sample of galaxies to shed light on these issues.  

One way to track the presence of AGN activity is through emission-line diagnostic diagrams, which identify the main ionizing mechanism in the interstellar gas by combining permitted and forbidden lines. Hydrogen permitted lines are due mainly to high-energy photons generated by young stars, thus serving as a proxy for star formation. The forbidden lines, mostly from ionised metals, originate mainly in collisionally excited states, which are usually related to the presence of AGN activity, traveling shocks, or the effect of evolved stars during a short, yet very hot and energetic phase known as post asymptotic giant branch \citep[post-AGB, e.g.][]{2019igfe.book.....C}.  Several diagnostic diagram have been proposed in the literature \citep[e.g.][]{2010MNRAS.403.1036C, 2018ApJ...856..171Z}, among which the Baldwin, Phillips \& Terlevich \citep[BPT, ][]{1981PASP...93....5B} diagram is the most commonly used.

Understanding the details of how AGN affect their host galaxies requires large samples to be able to account separately for the external environment  (e.g., mass and richness of host halo, local density) and internal properties (e.g., stellar mass $M_\text{stellar}$, galaxy velocity dispersion, morphology). The Sloan Digital Sky Survey (SDSS) provides optical imaging and spectroscopic data covering 14,555 squared degrees on the sky, providing a suitably large galaxy sample in the local ($z \leq 0.1$) Universe. 
Environmental information is available for SDSS galaxies \citep[e.g.][]{2007ApJ...671..153Y, 2014MNRAS.439..611W}, providing the necessary data for extensive studies of the effects of the environment on the properties of these galaxies \citep[e.g.][]{2014MNRAS.445.1977L, 2019MNRAS.484.1702P}. 

With such wealth of data we will be able to investigate the demographics of optically-identified AGN in both isolated and group/cluster central galaxies. We will study the interplay between internal and environmental properties driving AGN and fueling. To this end, we will first separate galaxies according to their environment and use the BPT diagram to investigate the fraction of strong AGN (Seyfert galaxies) in different environments. Additional factors such as galaxy mass, morphology, central velocity dispersion, and host halo mass, will be considered and linked, together with the environmental information, to the AGN and star-formation activity of the galaxies. Our aim is to identify the physical processes fueling AGN and driving star-formation activity, and causing both star-formation quenching and morphological transition. 

This paper is organized as follows: in \S2 we present the galaxy sample properties; in \S3 we separate galaxies according to their environment and stellar mass and compare their distribution on the BPT diagram; in \S4 we investigate the prevalence of strong AGN as a function of galaxy morphology and environment; in \S5 we directly trace the interplay between environment, AGN activity, and the evolutionary state of central galaxies through the star-formation rate versus stellar mass diagram; in \S6 we present the main results and our final conclusions. Throughout this paper we adopt a flat $\rm \Lambda CDM$ cosmology with $[\Omega_\text{M},\Omega_{\Lambda}, H_{0}] = [0.27,0.73,72\,$km$\,\rm s^{-1} Mpc^{-1}]$.
\renewcommand{\arraystretch}{1.5}

\section{Data and sample selection}
\label{sec:Sample_Data}

We use the single-fibre spectroscopy from the Sloan Digital Sky Survey Sixteenth Data Release (SDSS--DR16, \citealt{2009ApJS..182..543A}). We limit our sample to the redshift range $ 0.03 \leq z \leq 0.1$; the lower limit minimises aperture biases in the measured galaxy properties due to the fixed 3\,arcsec fibre diameter. To guarantee completeness, we only select galaxies with apparent Petrosian magnitude in the $r$-band brighter than 17.78, which corresponds to the spectroscopic completeness limit of the survey at $z = 0.1$. We also impose a signal-to-noise ratio (S/N)\footnote{Median signal-to-noise ratio of all good pixels in the galaxy spectrum.} threshold of 10 to ensure adequate spectral quality.

\subsection{Central galaxy selection and environment characterization}

From these data, we select central galaxies using the Yang catalogue \citep{2007ApJ...671..153Y}. This catalogue was built by applying a halo finder algorithm \citep{2005MNRAS.356.1293Y} to the New York University Value Added Galaxy Catalogue \citep{2005AJ....129.2562B}. Galaxies in the same halo are considered to be part of the same group/cluster. This allows us to define different environments for central galaxies: 1) isolated centrals -- galaxies living in halos occupied by a single galaxy ($N_\text{members} = 1$); 2) binary system centrals -- the most massive galaxy of a pair hosted in the same halo ($N_\text{members}=2$); 3) group centrals -- central galaxies in halos occupied by at least 3 galaxies, but no more than 10 ($3 \leq N_\text{members} < 10$); and 4) cluster centrals -- the central system in halos populated by more than 10 member galaxies ($N_\text{members}>10$). For halos hosting more than one galaxy, we define the central system to be the most massive one, with stellar mass calculated via equation (2) in \cite{2007ApJ...671..153Y}. We tested how the sample would change if we selected the brightest cluster galaxy (BCG) instead; we find no significant differences, since 98.2\% of the most massive galaxies are also BCGs. We further guarantee completeness regarding the halo mass estimate by imposing the conservative threshold presented in \cite{2009ApJ...695..900Y}, 
\begin{equation}
     \log(M_\text{halo}/M_{\odot}) \geq 12 + \frac{z_\text{c} - 0.085}{0.069},
    \label{eq:yang_mass_completeness}
\end{equation}
where $z_\text{c}$ denotes the median redshift of a given system (halo), and $M_\text{halo}$ is the halos mass estimated by \cite{2007ApJ...671..153Y}. 

\begin{figure}
    \centering
    \includegraphics[width = \columnwidth]{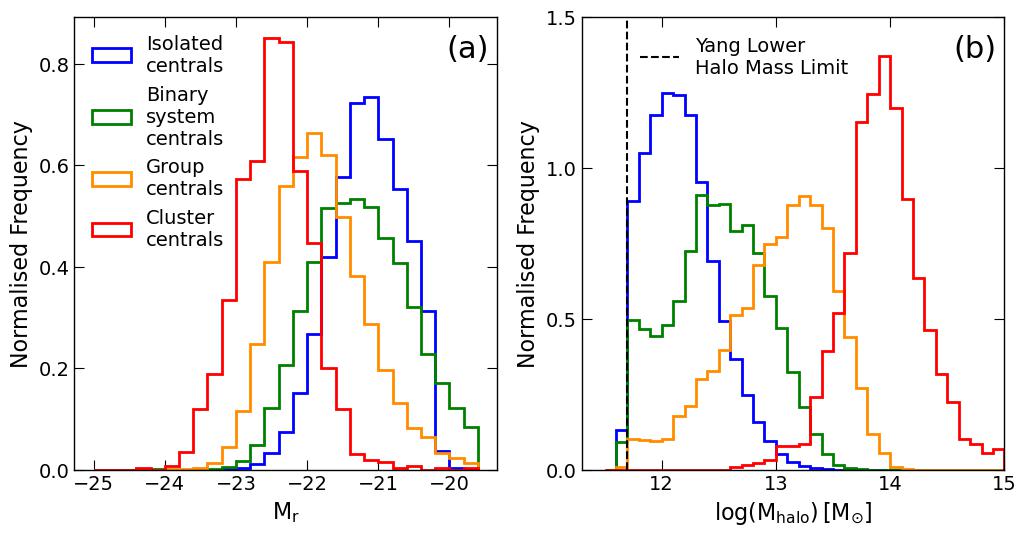}
   \caption{Distribution of the $r$-band absolute magnitude $M_\text{r}$ (left) and host-halo mass 
   $M_\text{halo}$ (right) for the four sub-samples of central galaxies: isolated centrals (blue line), centrals in binary system (green), in groups (orange), and in clusters (red). The vertical dashed line in the right panel show the lower halo-mass completeness limit from the Yang Catalogue.}
    \label{fig:Yang_Sample_Properties}
\end{figure}

In our analysis we will need to take into account that, as a result of separating central galaxies according to the number of satellites in their host halo, each sub-sample has different magnitude (and therefore stellar mass), and host halo-mass distributions, as shown in Fig.~\ref{fig:Yang_Sample_Properties}. 

The four sub-samples of central galaxies we have defined contain 59,605 isolated centrals, 11,639 in binary systems, 7,586 in groups, and 1,291 in clusters.

\subsection{Galaxy properties}

\subsubsection{Stellar masses, star-formation rates, and velocity dispersions}
\begin{figure}
   \centering
   \includegraphics[width = \columnwidth]{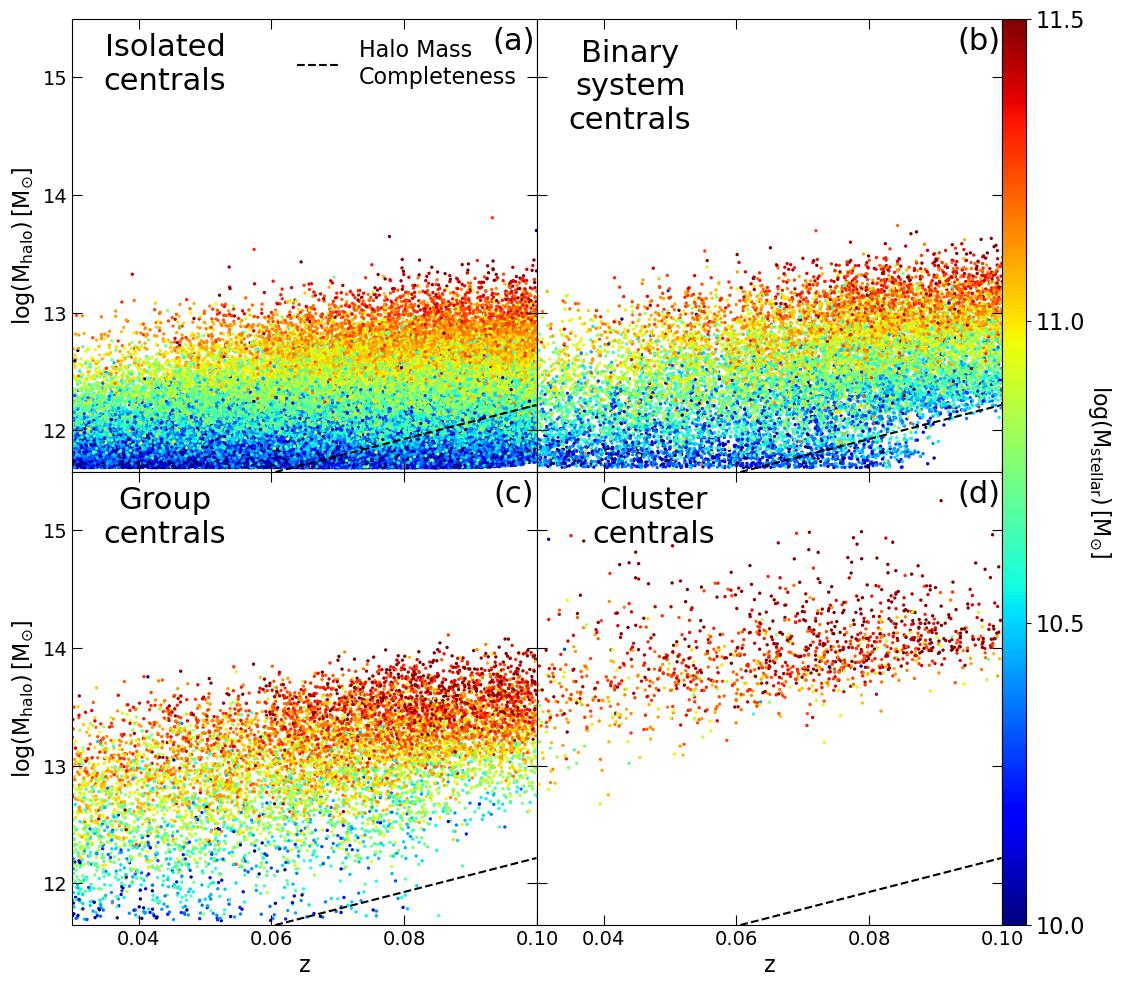}
  \caption{Relation between host-halo mass and redshift. Galaxies are separated into isolated centrals (top left), centrals in binary system (top right), group centrals (bottom left), and cluster centrals (bottom right). The dashed line in each panel indicates the halo mass completeness limit as a function of redshift (Equation~\ref{eq:yang_mass_completeness}). Points are coloured according to the galaxies' stellar mass.}
   \label{fig:Yang_Completeness}
\end{figure}

\begin{figure*}
    \centering
    \includegraphics[width = \textwidth]{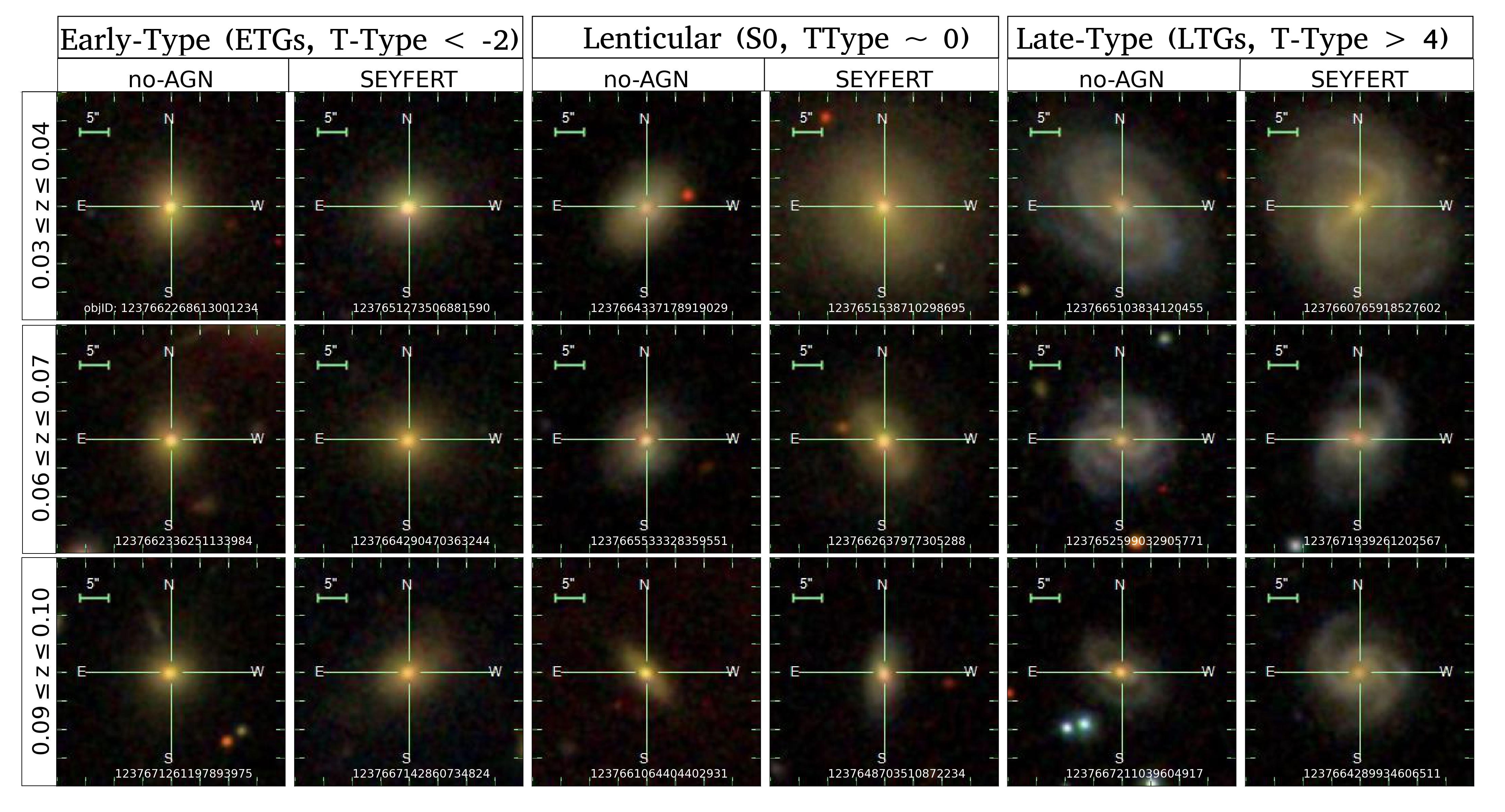}
   \caption{Comparison of the SDSS images of pairs of galaxies with different morphologies (T-Types) in three redshift bins (rows) for no-AGN (left image in each pair) and AGN (Seyfert, right image) galaxies classified using the BPT diagram.  It is apparent that the presence of an AGN doesn't alter the morphology of the host galaxy significantly. }
    \label{fig:TType_Mosaic}
\end{figure*}

The galaxies' stellar mass ($M_\text{stellar}$), star formation rate (SFR), and velocity dispersion ($\sigma_\text{galaxy}$) are obtained from the Max Planck Institut f\"ur Astrophysik -- John Hopkins University catalogue \citep[MPA--JHU, ][]{2004ApJ...613..898T, 2004MNRAS.351.1151B}. It provides reliable measurements and their associated errors for galaxies in the SDSS--DR8 without spectral anomalies. An important feature is the way they estimate the galaxies' SFR: it is first derived using the $\rm H\alpha$ emission-line flux for galaxies classified as `star-forming' in the BPT diagram (see Section~\ref{sec:BPT_Class}), and then extrapolated to other galaxies using a calibration between SFR and the 4000\AA\ break. This is particularly important in our study since this method prevents the SFR estimates being contaminated by possible AGN contribution to the $\rm H\alpha$ flux. An indication of the typical uncertainties in the galaxies' properties is provided in Table~\ref{table:spectra_errors},

\begin{table*}
\caption{Median uncertainties in the $M_\text{stellar}$, SFR, and $\sigma_\text{galaxy}$ estimates for galaxies in different stellar-mass ranges. For the stellar masses and SFRs we calculate the relevant $1$-$\sigma$ uncertainty $\Delta(1\sigma)$ for each galaxy from the $\rm 86^{th}$ and $\rm 14^{th}$ percentiles $Q_{86}$ and  $Q_{14}$ using $\Delta(1\sigma) = 0.5 \times (Q_{86} - Q_{14})$. The velocity dispersion uncertainties are provided by the MPA--JHU catalogues. } 

\label{table:spectra_errors}
\resizebox{0.7\textwidth}{!}{%
\begin{tabular}{lccc}
\hline
     & $ 10 \leq \log(M_\text{stellar}/M_{\odot}) < 10.5$ & $10.5 \leq \log(M_\text{stellar}/M_{\odot}) < 11$ & $11 \leq \log(M_\text{stellar}/M_{\odot})$ \\ \hline
$\Delta\log(M_\text{stellar})$ [$M_\odot$] & 0.09                                            & 0.09                                            & 0.09                                     \\ 
$\Delta\log(\text{SFR})$ [$M_\odot/yr$]    & 0.31                                            & 0.99                                            & 1.02                                     \\

$\Delta\sigma_\text{galaxy}$ [km/s]& 11.85                                            & 10.66                                            & 11.44                                     \\ \hline
\end{tabular}%
}
\end{table*}

The galaxies in our sample have stellar masses in the  range $10^{10}M_\odot \le M_\text{stellar} \le 10^{11.5} M_\odot$. However, Equation~\ref{eq:yang_mass_completeness} implies that for some of the sub-samples we lose some of the less massive systems at high redshift (see Fig.~\ref{fig:Yang_Completeness}). This affects mostly isolated centrals, for which we only reliably 
cover the full stellar mass range up to  $z\sim0.06$. Since we only consider a relatively narrow redshift range, we do not expect significant evolution with look-back time, and therefore the fact that for some low-mass galaxies we don't sample the full redshift range should not affect our results.  

\begin{figure*}
    \centering
    \includegraphics[width = \textwidth]{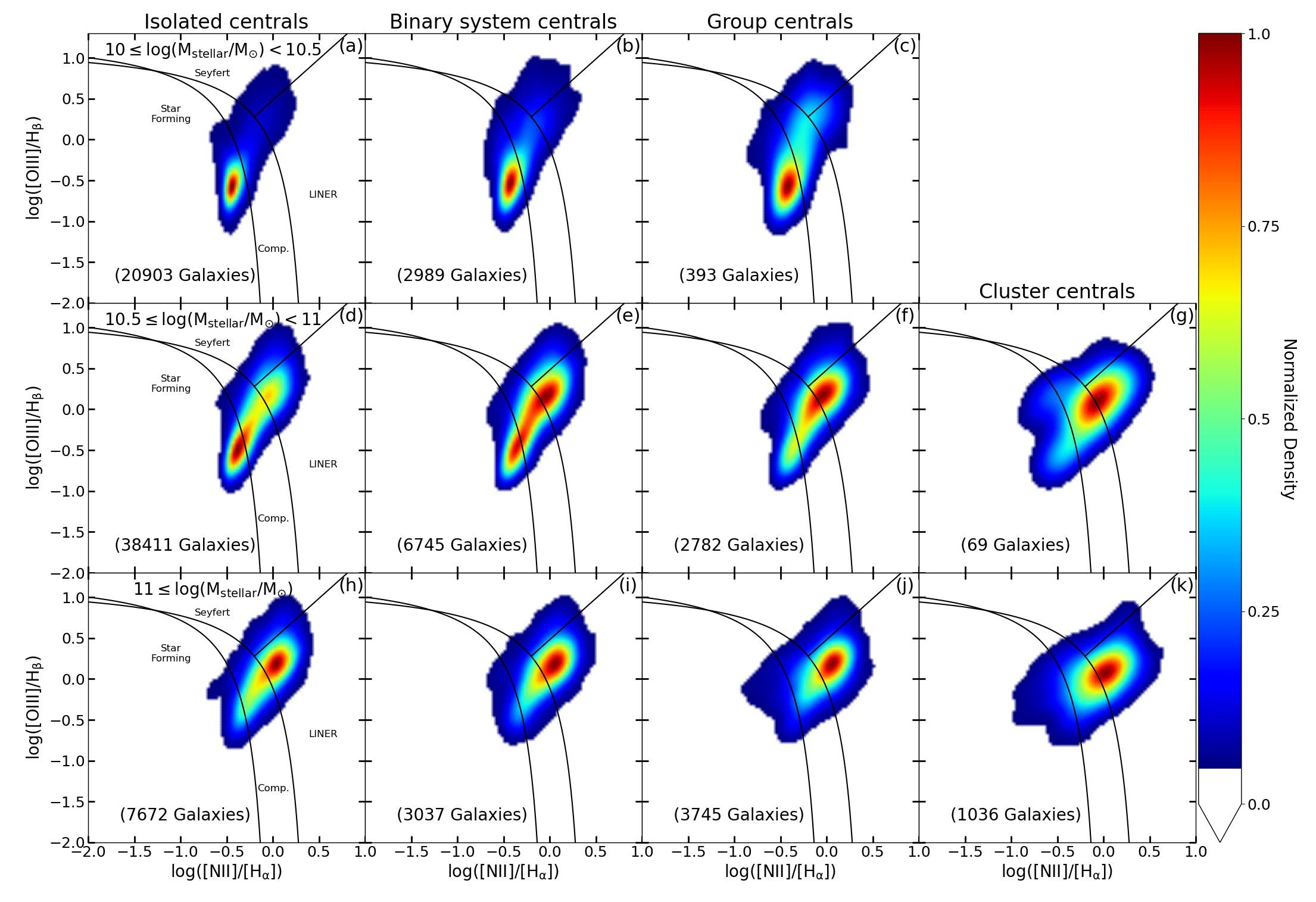}
    \caption{Normalized Gaussian-kernel-smoothed distribution of central galaxies in the BPT diagram for different environments (from left to right, and in order of `environmental richness', isolated centrals, centrals in binary system, in groups, and in clusters) and in three mass bins (from top to bottom, $10^{10} \leq M_\text{stellar}/M_{\odot} < 10^{10.5}$; $10^{10.5} \leq M_\text{stellar}/M_{\odot} < 10^{11}$; and $M_\text{stellar}/M_{\odot} \geq 10^{11}$). The lines separating different regions of the diagram are described in section~\ref{sec:BPT_Class}. The number of galaxies in each sub-sample is indicated towards the bottom left of each panel. Note that our sample does not contain cluster centrals with log$(M_\text{stellar}/M_{\odot}) < 10.5$.} 
    \label{fig:BPT_Smooth}
\end{figure*}

\subsubsection{BPT classification}
\label{sec:BPT_Class}

The BPT diagram uses the flux ratios of two pairs of emission lines,  $\rm [OIII]_{\lambda 5007}$ over $\rm H\beta$, and $\rm [NII]_{\lambda 6584}$ over $\rm H\alpha$ \citep{1981PASP...93....5B}. The lines in each pair are close in wavelength, minimising extinction effects. 
The MPA--JHU catalogues provide the required emission-line fluxes. These are measured after the continuum and stellar absorption lines have been subtracted\footnote{The stellar contribution to the spectra is estimated through spectral fitting using single stellar population models from \cite{2003MNRAS.344.1000B}. See  \cite{2004MNRAS.351.1151B} for more details}.  Table \ref{table:line_errors} shows the median uncertainties in the emission-line flux measurements.

On a BPT diagram (see Fig.~\ref{fig:BPT_Smooth}), the line demarcating the region where the dominant ionisation mechanism is star formation can be  derived theoretically as the upper limit for the line ratios that can arise in gas ionised by an starburst \citep{2001ApJ...556..121K}.  A second line is defined observationally by \cite{2003MNRAS.346.1055K} separating systems whose interstellar gas is mainly collisionally ionized. The intermediate region enclosed by these two lines contains galaxies with `composite' spectra, such that their interstellar gas is ionized by a combination of both radiative and collisional mechanisms. Galaxies above and to the right of the \cite{2001ApJ...556..121K} line are further divided into two different classes,  Seyferts and LINERs (Low Ionization Nuclear Emission Regions), using another line defined by \cite{2001ApJ...556..121K}. Fig.~\ref{fig:BPT_Smooth} presents BPT diagrams for several galaxy sub-samples, showing the lines dividing the different regions.

Using emission-line data, the MPA--JHU catalogue classifies galaxies according to their dominant ionising mechanism using the BPT diagram with the dividing lines defined above.  Galaxies are thus divided into six categories:  1) star forming, 2) low S/N star forming, 3) composite, 4) Seyfert, 5) LINER, and 6) unclassified. This last class comprises galaxies with no reliable classification in the BPT diagram due to at least one of the four relevant emission lines having S/N smaller than 3. Furthermore, in order to increase the robustness of our results, we only accept the BPT classification for galaxies with $\rm H\alpha$ equivalent width ($EW(\text{H}\alpha$)) greater than 3\AA, which is the threshold used in the WHAN diagram to separate LINERs and Retired galaxies \citep{2010MNRAS.403.1036C}. The flux of weaker lines would be highly uncertain in the presence of underlying stellar absorption lines even after correcting for them. This way we avoid contaminating our sample with weak emission line galaxies that can be explained by the presence of old stellar population. Hereafter we will refer to the joint sample of `unclassified' galaxies and those with  $EW(\text{H}\alpha) \leq 3$\AA\ as `Passive+Retired' (P+R) galaxies. As we will show below, these Passive+Retired galaxies have negligible star formation, as expected.

It is important  to point out that, while the emission lines of Seyfert galaxies are well explained by the presence of a strong AGN, there is considerable uncertainty -- and some controversy -- on the origin of the emission lines observed in LINERs. However, it is now generally accepted that, in addition to low-activity AGN, LINER emission may also be due to the presence of an old stellar population \citep{2010MNRAS.403.1036C} containing very hot low-mass stellar relics such as post-AGB stars. Since the object of this work is to investigate the effect of AGN activity on their host galaxies, we will not consider either composite-spectra galaxies, since we want to make the cleanest possible separation between star-forming and Seyfert galaxies, or LINERs, because it is very likely that these objects are not driven primarily by a central SMBH engine, given that we use
single fibre spectroscopy, so that we cannot probe the spatial distribution of the nebular emission. In what follows we will only consider galaxies in the Seyfert region of the BPT diagram as unequivocal optically-selected AGN.

\begin{table*}
\caption{Median uncertainties in line flux measurements for galaxies in different stellar-mass ranges. The flux units are $10^{-17}\,{\rm erg\,s^{-1}cm^{-2}}$.}
\label{table:line_errors}
\resizebox{0.7\textwidth}{!}{%
\begin{tabular}{lccc}
\hline
 & $\rm 10 \leq log(M_{stellar}/M_{\odot}) < 10.5$ & $\rm 10.5 \leq log(M_{stellar}/M_{\odot}) < 11$ & $\rm 11 \leq log(M_{stellar}/M_{\odot})$ \\ \hline
$\Delta\rm [NII]_{\lambda 6584}$  & 2.81                                            & 3.57             & 5.17                                     \\
$\Delta\rm H\alpha$          & 3.53                                            & 3.86                                            & 5.59                                     \\
$\Delta\rm [OIII]_{\lambda 5007}$ & 2.62                                            & 3.32                                            & 4.45                                     \\
$\Delta\rm H\beta$           & 2.61                                            & 3.05                                            & 4.24          \\ \hline
\end{tabular}%
}
\end{table*}

\subsubsection{Morphologies}

When the star-formation properties of galaxies change, their morphologies may also change. Here we use the T-Type parameter, first introduced by \cite{1963ApJS....8...31D}, to describe a galaxy's morphology. To aid our analysis it is convenient to use a revised T-Type parameter which varies continuously from $-3$ to 6 \citep{2015MNRAS.446.3943M} instead of the original discrete values.  As usual, a T-Type value $\leq0$ indicates an early-type galaxy, while T-Type values $>0$ correspond to late-type morphologies. We adopt the T-Type values from the work of \cite{2018MNRAS.476.3661D}, which was based on the application of a deep-learning convolutional neural network algorithm to~670,722 galaxies from the SDSS-DR7 database.  

It is conceivable that the presence of a strong AGN may influence the morphological classification of the host galaxy. To ensure that such effect is not significant enough to alter the conclusions of this work, we examine the images of a subset of randomly selected galaxies classified as Seyfert and compare them with the images of non-Seyfert galaxies with similar morphologies. Fig.~\ref{fig:TType_Mosaic} shows this comparison for a sample of galaxies, Seyfert and non-Seyfert, for different morphologies and redshifts. Visual inspection indicates that that the presence of an AGN does not alter the morphology significantly, at least at optical wavelengths in these nearby galaxies. 

Furthermore, in order to check that the results of our analysis do not depend critically on the specific choice of morphological data, we repeated our analysis using independent visual morphologies from the the Nair and Abraham catalogue \citep{2010ApJS..186..427N}. These morphologies do not rely on automated algorithms, being based on human inspection of the galaxy images. Notwithstanding the different sample sizes of both morphological catalogues, the results obtained are completely compatible. 

The results of both tests indicate that the morphological data we adopt are suitable for this analysis. 

\section{Star formation and AGN activity as a function of stellar mass and environment}
\label{sec:SF_AGN_stellar_mass_environment}
Fig.~\ref{fig:BPT_Smooth} shows the distribution of central galaxies in the BPT diagram for the different environments we are considering (isolated centrals, centrals in binary system, in groups, and in clusters), in three stellar-mass bins ($10^{10} \leq M_\text{stellar}/M_{\odot} < 10^{10.5}$; $10^{10.5} \leq M_\text{stellar}/M_{\odot} < 10^{11}$; and $M_\text{stellar}/M_{\odot} \geq 10^{11}$).

We observe a decreasing fraction of star-forming galaxies for increasing stellar mass, irrespective of environment. This is in agreement with the so-called `downsizing' scenario, and suggests that, as star formation decreases, other ionizing mechanisms become more relevant and prominent at high stellar masses. In particular, the increasing fraction of massive galaxies where collisionally driven ionisation dominates suggests that AGN activity may be involved in quenching star formation \citep{2017FrASS...4...10C, 2020MNRAS.499..230B}. Within a given stellar mass bin, we find a decreasing fraction of star-forming galaxies as `environmental richness' increases. For instance, in the low stellar-mass bin the fraction of star-forming galaxies decreases from 31\% in isolated centrals to 17\% in cluster centrals.

\begin{figure}
    \centering
    \includegraphics[width = \columnwidth]{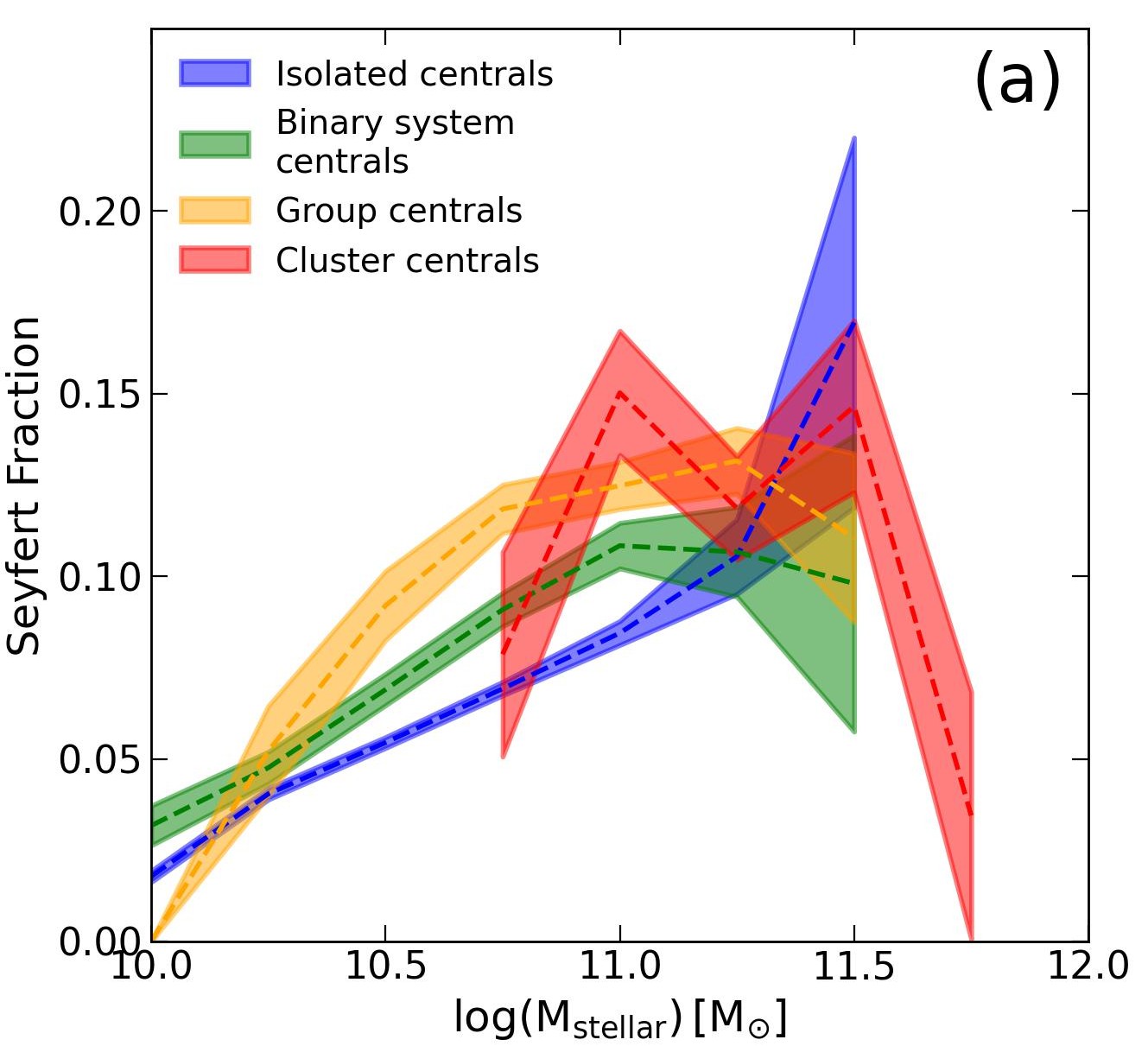}
    \caption{Seyfert fraction as a function of stellar mass, separating galaxies according to environment. Dashed lines and shaded areas represent medians and 1-sigma uncertainties, respectively.}
    \label{fig:seyfert_overall_fraction}
\end{figure}

Moreover, Fig.~\ref{fig:BPT_Smooth} indicates that AGN presence also depends on environment. For instance, for low stellar mass centrals there is a higher Seyfert fraction in group centrals (11\%), compared with 6\%, 8\% and 7\% for isolated, binary system and cluster centrals, respectively. Although AGN are always a small minority in each category, the change is sizeable. In more detail, Fig.~\ref{fig:seyfert_overall_fraction} shows the Seyfert fraction as a function of stellar mass, separating galaxies according to their environment. We calculate the fraction in bins of 0.25 dex, from $\log (M_{\rm stellar}/M_\odot)=10$ to~12, limiting the calculation to bins with at least 20 galaxies. We interpolate linearly between neighbouring bins. Here and in subsequent figures we estimated the uncertainties using a standard bootstrap technique with 1000 re-samplings. We find that the Seyfert fraction increases with stellar mass for isolated, binary system and group centrals, whereas cluster centrals show an almost constant fraction ($\sim 12.5 \pm 4 \%$ from $\rm 10^{11}$ to $\rm 10^{11.5}M_{\odot}$ followed by a steep decrease towards $\rm 10^{11.75}M_{\odot}$ ($\rm 3 \pm 6 \%$ in the last bin).  
Comparison between different environments shows an increasing Seyfert fraction for increasing richness in the $\rm 10^{10.25}$ and $\rm 10^{11.25}M_{\odot}$ range. This suggests there may be a significant environmental effect on the fueling of AGN activity, which supports that in order to understand AGN feedback on central galaxies we need to analyse the complex interplay between internal and external factors in detail.

\section{The demographics of AGN as a function of the internal and environmental properties of central galaxies}
\label{sec:demographics}

AGN activity depends on the availability of gas that can be accreted onto the central SMBH. It is generally accepted that most galaxies in the mass range we are considering contain a SMBH at their centre.  The fact that only a small fraction of galaxies contain a strong AGN (e.g., Seyfert galaxies) can be attributed to the large amounts of gas required to fuel strong AGN  \citep{2014MNRAS.437.1199C, 2016MNRAS.458L..34E}. By tracing the fraction of Seyferts as a function of the galaxies' properties and environment we may be able to shed light on the combination of internal and environmental factors needed to drive AGN fueling in central galaxies, and consequently investigate the effect of AGN activity on galaxy evolution.

\subsection{The relation between AGN and the morphology of their host galaxies}
\label{sec:AGN_morphology}

\begin{figure*}
    \centering
    \includegraphics[width = \textwidth]{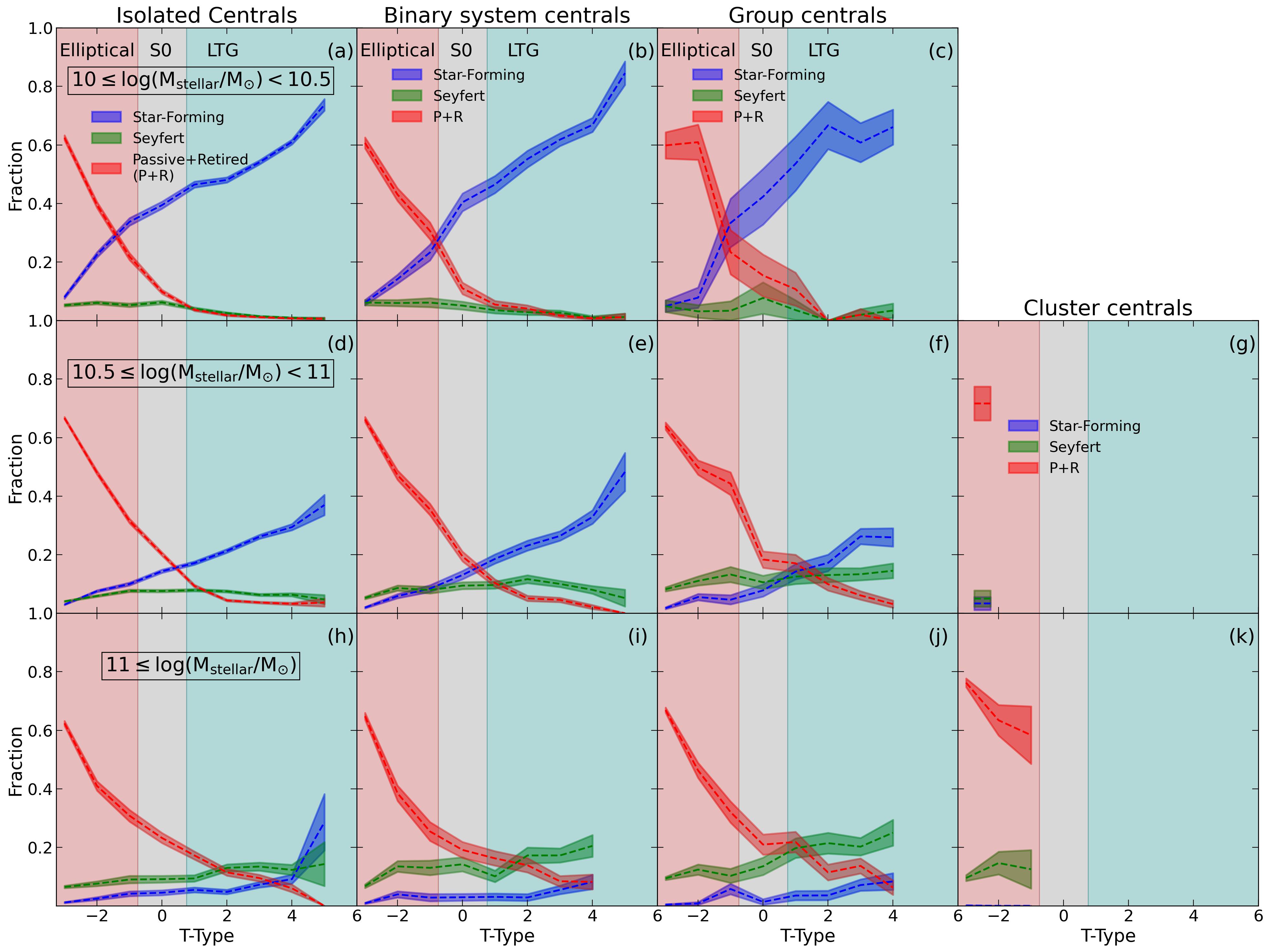}
    \caption{Fraction of star-forming (blue), Seyfert (green) and P+R (red) galaxies as a function of morphological T-Type. As in Fig.~\ref{fig:BPT_Smooth}, galaxies are divided according to their environment (columns) and stellar mass (rows). The red, gray and aqua colored backgrounds correspond to  the T-Type ranges of ellipticals, lenticulars, and spirals, respectively. Median error bars are shown on centre right of each panel. }
    \label{fig:Fract_vs_TType}
\end{figure*}

In Fig.~\ref{fig:Fract_vs_TType} we show the fraction of Seyferts as a function of the T-Type of their host galaxies. To provide context, we also show the fraction of star-forming and P+R galaxies. To guarantee reliable results, we also impose a minimum of 20 galaxies per bin to calculate the relevant fractions. For reference, we define three T-Type ranges: $-3 \le \text{T} <  -0.75$ corresponds to elliptical galaxies; $-0.75 \le \text{T} \le 0.75$ to lenticular (S0) galaxies; and $0.75 < \text{T} \le 6$ to spiral galaxies. These limits were chosen to be consistent with the uncertainties in T-Type ($\Delta $T-Type$\,\sim0.5$); changes in the bin boundaries of this size will not significantly change our results. Elliptical and lenticulars are often combined under the label `early-type galaxies' (ETGs), while spirals are `late-type galaxies' (LTGs). Overall, the decreasing fraction of star-forming systems for decreasing T-type values (from LTGs to ETGs) simply reflects the fact that most LTGs host significant star formation, whereas the opposite trend observed for P+R galaxies indicates that most ETGs do not form many stars. Although this basic picture is true for all galaxy masses and environments, the actual fraction of galaxy type changes both with mass and environment. 

The fraction of Seyfert galaxies is always relatively small, increasing with the galaxies' stellar mass in all environments. In agreement with the results of previous works, we find that low-mass galaxies are less likely to host an AGN than massive ones \citep{2008MNRAS.391..785W}. This can be interpreted as an indication that the fueling of an AGN depends on the amount of gas retained by the galaxy's host halo, which correlates with the halo mass. For the two lowest stellar mass bins, we find an almost constant fraction of Seyferts for morphologies in all environments, although the small numbers make this invariance somewhat uncertain. This low fraction may indicate that, regardless of their morphology, these systems are rarely able to hold a large enough gas reservoir to both fuel star formation and AGN activity. In this case, as soon any AGN feedback is triggered, all the neutral hydrogen gas is removed, preventing any subsequent AGN activity \citep{2022ApJ...933L..12G}.

For high stellar masses, the Seyfert fraction is significantly higher than in low-mass galaxies. Quantitatively, the average over all environments for the lower stellar mass bin gives a Seyfert fraction of $\rm 5 \pm 3\%$, in comparison with $\rm 13 \pm 4\%$. In the most massive bin, Seyferts are more abundant than star-forming galaxies in all environments for virtually all morphologies. There are more than twice as many high-mass galaxies classified as Seyferts than as star-forming even for late-type spirals.  This could simply suggest that, in the absence of star formation as an ionising mechanism, even a relatively modest level of AGN activity becomes apparent. But it could also mean that AGN activity is linked to (and perhaps the cause of) star-formation suppression in massive galaxies. 

Furthermore, the decreasing fraction of Seyferts when we transition from spirals to ETGs provides some support to the idea that AGN feedback plays some role in the morphological transition process, and in generating the morphological diversity we observe in present-day galaxies \citep{2009MNRAS.399.2172R, 2016MNRAS.463.3948D, 2022arXiv220909270V}. 
Our analysis further shows that the slope of the relation between AGN activity and morphology depends on environment, increasing with `environmental richness': it is flatter for isolated centrals and for those in binary systems than for group and cluster centrals. The numerical values of the slopes of the relation between Seyfert fraction and T-Type in Fig.~\ref{fig:Fract_vs_TType} are $0.009 \pm 0.003$, $0.015 \pm 0.009$, $0.022 \pm 0.010$ and $0.042 \pm 0.009$ for isolated, binary, group, and cluster centrals respectively, showing that environment plays a role. 

\subsection{The relation between AGN, halo mass, and central velocity dispersion of their host galaxies}
\label{sec:halo_mass_dispersion}
\begin{figure}
    \centering
    \includegraphics[width = \columnwidth]{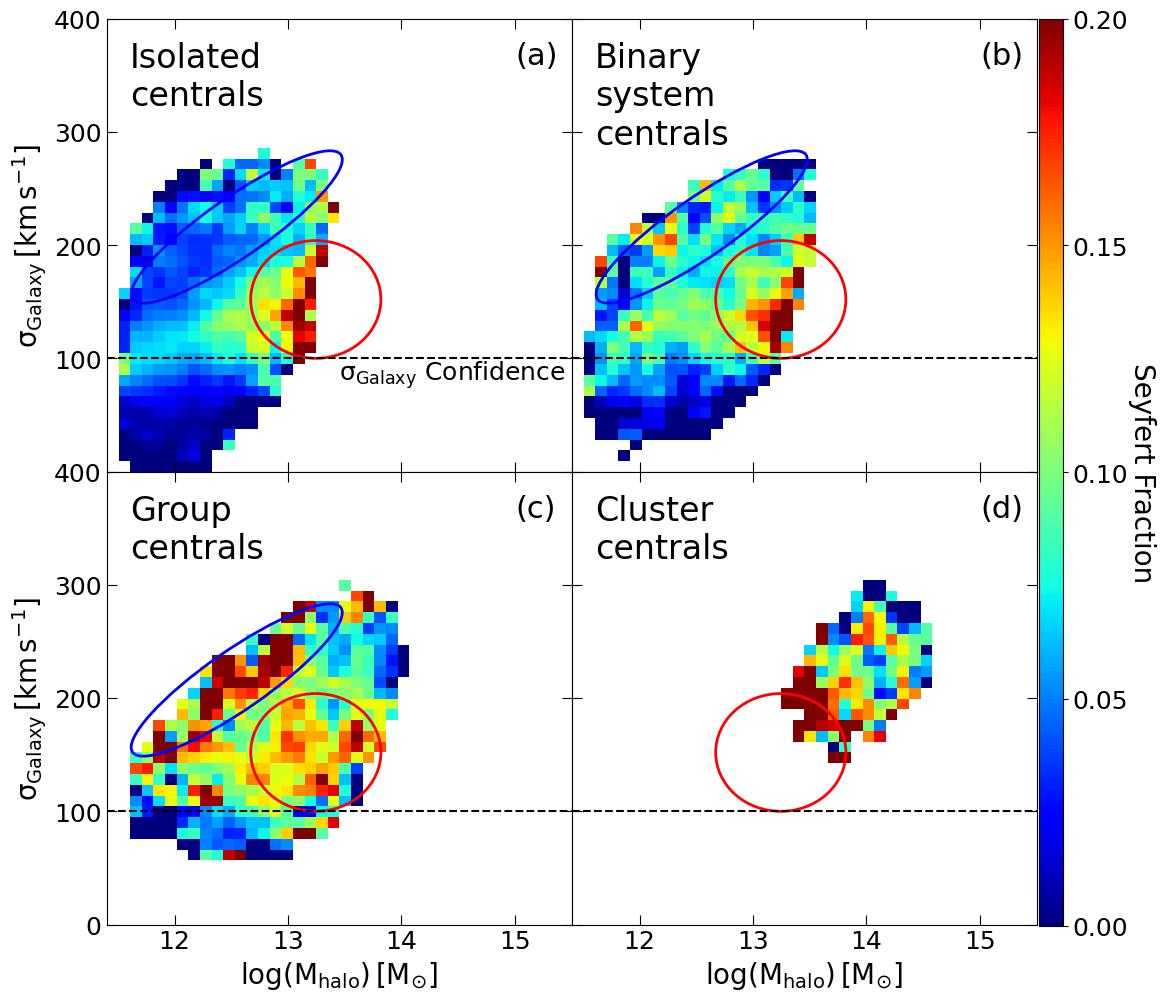}
    \caption{Seyfert fraction in the galaxies' central velocity dispersion ($\sigma_\text{galaxy}$) versus Halo Mass ($M_\text{halo}$) diagram. We separate central galaxies according to their environment into isolated centrals (top left), centrals in binary system (top right), in groups (bottom left) and in clusters (bottom right). The black dashed line shows the limit below which $\sigma_\text{galaxy}$ values from the MPA--JHU catalogue become unreliable. Note that galaxy stellar mass correlates reasonably well with $M_\text{halo}$. Circles and ellipses indicate regions of special interest (see text for details). 
    }
    \label{fig:Fraction_Sigma_vs_Mhalo}
\end{figure}

The availability of gas and the mass of the central SMBH are both key factors in AGN activity. The total gas mass is related, albeit in a complex way, to the mass of the host galaxy halo ($M_\text{halo}$), while the mass of the SMBH correlates with the central velocity dispersion of the galaxy ($\sigma_\text{galaxy}$; \citealt{2002ApJ...574..740T,2009ApJ...698..198G}). To bring all these parameters together, we show in Fig.~\ref{fig:Fraction_Sigma_vs_Mhalo} the median fraction of Seyferts in the $\sigma_\text{galaxy}$ vs.\ $M_\text{halo}$ plane for central galaxies in different environments. For each pixel, we derive the Seyfert fraction median and its uncertainty using a bootstrap technique, as before. We will limit our analysis to galaxies with $\sigma_\text{galaxy}>100\,$km/s, since below this value the galaxy velocity dispersions provided by the MPA--JHU catalogue become unreliable due to the resolution of SDSS spectra. In order to confirm the robustness of this trend, we repeated this analysis using 50\% larger bins to reduce the statistical uncertainties. None of the trends described below changed. Furthermore, we applied Gaussian smoothing to the density maps to enhance the visual appearance of the observed overdensities. This also confirmed our conclusions below. We decided to present the raw binned density maps without smoothing since they provide a fairer representation of the data and their uncertainties. 

Panel (a) of Fig.~\ref{fig:Fraction_Sigma_vs_Mhalo} shows that the largest Seyfert fraction in isolated centrals is found in galaxies residing in the largest mass halos ($M_\text{halo} \geq 10^{13} M_{\odot}$) and at intermediate velocity dispersions ($100 \leq \sigma_\text{galaxy} \leq 200\,$km\,s$^{-1}$). A similar locus with a relatively high fraction of Seyfert galaxies (highlighted by the red circles) is also found in the other panels. Quantitatively, this region has median Seyfert fractions of $\rm 17 \pm 9 \%$, $\rm 15 \pm 7\%$, and $\rm 14 \pm 3\%$ in isolated, binary system, and group centrals respectively. A high $M_\text{halo}$ suggests a large reservoir of gas \citep{2014ApJ...795..144C,2019MNRAS.485.3783D}, which may be used to fuel the AGN. However, a direct relation between the size of the gas reservoir and AGN fueling still needs to be verified by observations. The intermediate velocity dispersion range occupied by these galaxies corresponds to the transition region between late- and early-type morphologies, providing additional circumstantial evidence of the effect strong AGN may have on the galaxies' morphological transition.

Panel (c) of Fig.~\ref{fig:Fraction_Sigma_vs_Mhalo} shows, for group centrals, a second region with high Seyfert fraction (indicated by the blue ellipse). These are galaxies with relatively high velocity dispersion for their halo masses. The fraction of Seyferts in the same location of the diagram decreases systematically when we move to centrals in binary systems and isolated ones, with median values of $2 \pm 2\%$, $ 9 \pm 4\%$ and $18 \pm 7\%$ for isolated, binary system and group centrals, respectively. Clearly, in this region of the diagram, the environment affects AGN fuelling. For group centrals, at a given halo mass Seyferts are mainly found in galaxies with high $\sigma_\text{galaxy}$, indicating that, even if the gas reservoir is the same, a larger SMBH mass increases the probability of AGN activity. These high velocity dispersion galaxies tend to be ETGs. That this happens mostly in the group environment may suggest that group centrals are more likely to have experienced a recent interaction with other galaxies (their satellites) than isolated centrals and centrals in binary systems. Such interactions can drive gas to the central system, boosting its fuel reservoir. In support of this, \cite{2014MNRAS.445.1977L} show that central galaxies in groups hosted a more recent star formation episode in comparison to their isolated counterparts, indicating a recent supply of gas. Additionally, \cite{2012A&A...548A..37K} show and increase in [OIII]5007\AA\ when moving closer to the central galaxy. Finally, \cite{2018PASJ...70S..31O} show an enhanced galaxy overdensity around quasar pairs, which are more common in more massive halos. Together with our results, these results provide evidence that the interaction between galaxies can enhance AGN activity. Nevertheless, a more direct test to support this scenario would be the observation of the gas component of a large sample of central galaxies covering a wide range of halo masses, with enough resolution to detect inflows and outflows.

Moving to cluster centrals, panel (d) of Fig.~\ref{fig:Fraction_Sigma_vs_Mhalo} shows the highest fraction of Seyferts at the low-halo-mass end of the distribution ($\rm 20 \pm 10\%$). The innermost regions of the most massive clusters are already ``fully developed'' \citep{1978ApJ...219...18B, 1980ApJ...241..521C, 1997ApJ...476L...7C, 2021MNRAS.503.3065S}, and the probability of satellite galaxies interacting with the cluster central -- and thus providing fuel -- is lowest for the most massive clusters due to the very high relative velocities of the satellite galaxies \citep{1943ApJ....97..255C}. 
\begin{figure}
    \centering
    \includegraphics[width = \columnwidth]{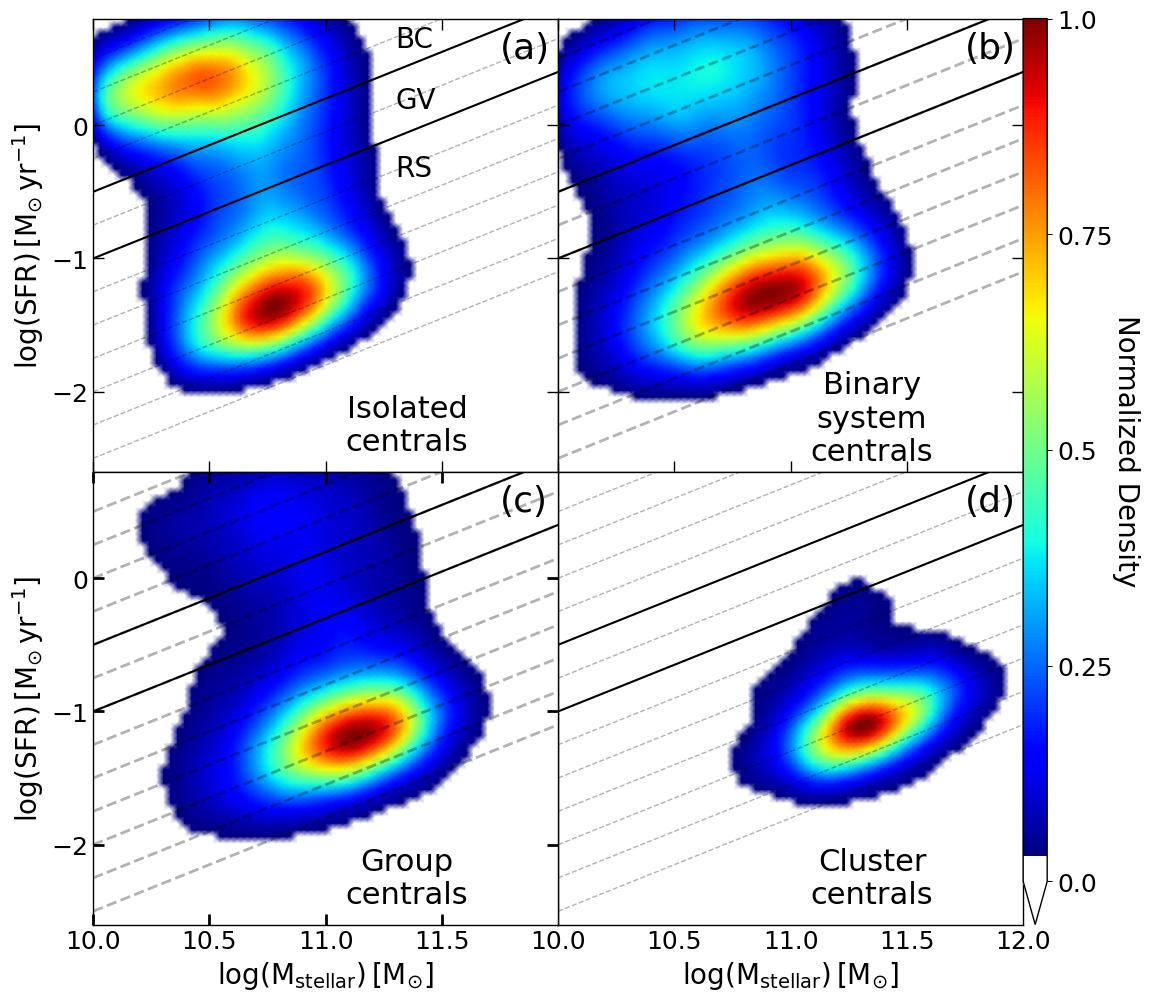}
   \caption{Gaussian kernel smoothed distribution of central galaxies in the star formation rate versus stellar mass plane. The separating lines between blue cloud, green valley, and red sequence regions are shown as yellow solid lines. The white dashed lines have the same slope as the yellow lines, but with varying intercept. Central galaxies are divided into isolated centrals (upper left), centrals  in binary systems (upper right), in groups (lower left), and in clusters (lower right).}
    \label{fig:Smooth_Density}
\end{figure}
From the AGN demographics presented in this section, we have found that, although stellar mass is the strongest statistical predictor of AGN activity (c.f.\ Sections~\ref{sec:SF_AGN_stellar_mass_environment} and~\ref{sec:AGN_morphology}), other factors such as morphology, environment, SMBH mass, and halos mass also play important roles. For massive galaxies, the relatively high AGN fraction in LTGs and the dependence of AGN frequency on T-Type suggests that feedback from the AGN is relevant in the morphological transformation of the galaxies. The details of the distribution of Seyfert fractions in the $M_\text{halo}$ versus $\sigma_\text{galaxy}$ plane provides further insight into how the environment may drive the fueling of AGN activity. For a given halo mass range, the high fraction of Seyferts in group centrals is found at high velocity dispersion, suggests that interactions with satellite group galaxies can enhance the gas reservoirs that fuel the AGN of group central galaxies, which are predominantly ellipticals. 

\begin{figure*}
    \centering
    \includegraphics[width = \textwidth]{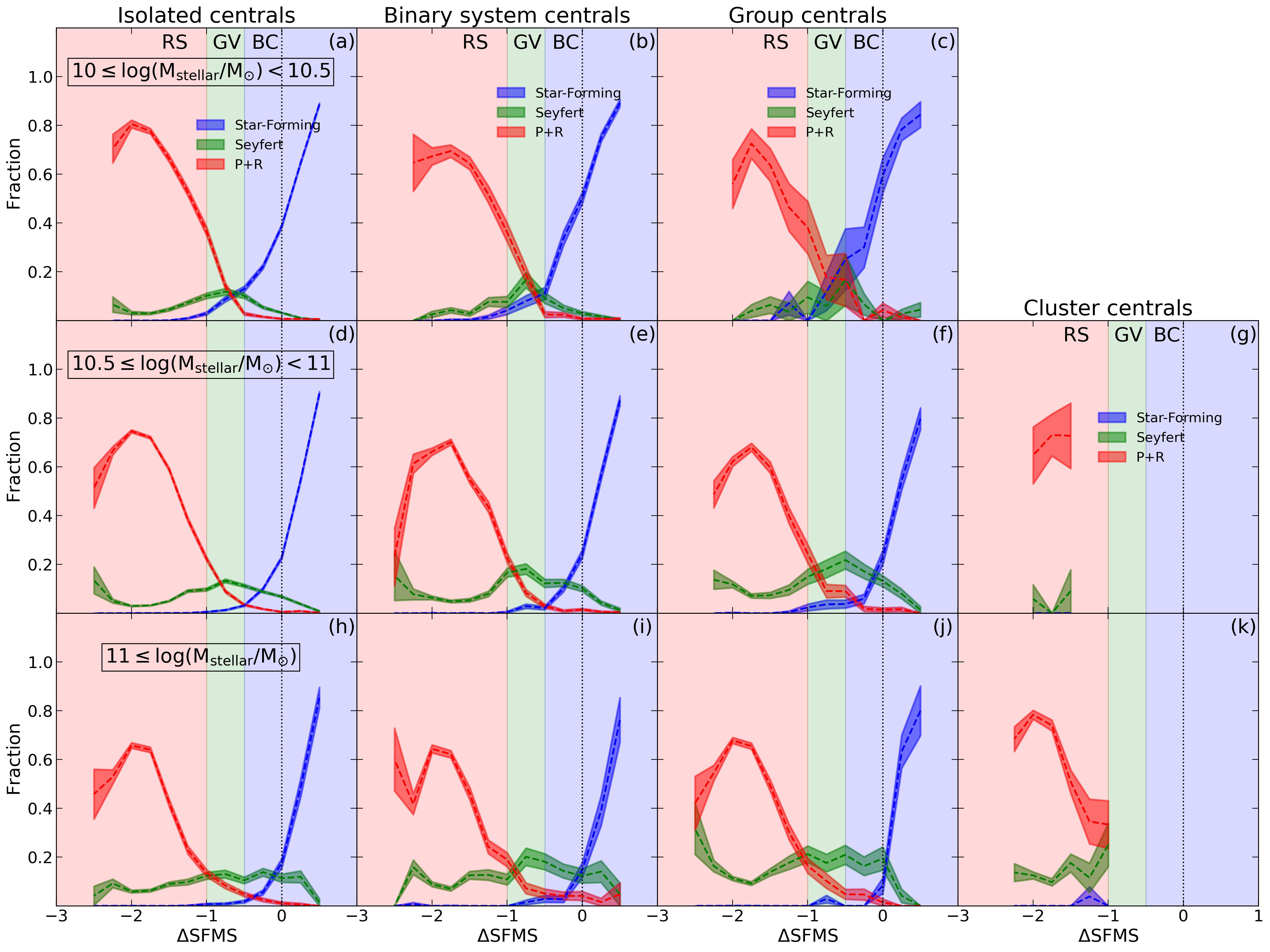}
   \caption{Fraction of central galaxies classified as star forming (blue), Seyferts (green) and P+R galaxies (red) as a function of the vertical distances $\rm \Delta SFMS$ from the star-formation main sequence (SFMS) for galaxies in different environments (columns) and stellar-mass ranges (rows). The blue, green, and red backgrounds represent the blue cloud, green valley, and red sequence regions, as defined in Fig.~\ref{fig:Smooth_Density}. 1-sigma uncertainties are presented as shaded regions. Positive $\rm \Delta SFMS$ values correspond to galaxies with higher star-formation activity than galaxies in the SFMS, while negative values indicate star-formation suppression relative to the SFMS.
   }
    \label{fig:Dissection_SFR_Mstellar_Fraction}
\end{figure*}

\section{The role of AGN on the star-formation activity of central galaxies}
\label{sec:AGN_SFR}

In this section we focus on how the interplay between AGN activity, stellar mass, and environment drives the evolution of the star-formation activity of central galaxies. We quantify the galaxies' star-formation activity through the the SFR vs.\ $M_\text{\rm stellar}$ diagram, shown in Fig.~\ref{fig:Smooth_Density} for central galaxies in different environments. The Blue Cloud (BC), Green Valley (GV), and Red Sequence (RS) regions are defined by two dividing lines: $\text{log(\rm SFR)} =  0.7 \log(M_\text{\rm stellar}) - 7.5$ separates the BC from the GV, and  $\text{log(\rm SFR)} =  0.7 \log(M_\text{\rm stellar}) - 8.0$ separates the GV from the RS \citep{2020MNRAS.491.5406T, 2022MNRAS.509..567S}. To avoid confusion, we refer to galaxies located in the BC  region of this diagram `Blue-Cloud galaxies', reserving the term `star-forming galaxies' for those classified as such in the BPT diagram. 

Consistent with the results of previous sections, we find that the fraction of galaxies in the BC, i.e. with significant star formation, decreases with stellar mass and environmental richness, while the RS shows the opposite trend.

In order to provide a more detailed continuous description of the transition from the BC to the RS, we divide the SFR vs.\ $M_\text{stellar}$ plane into 12 zones separated by the white dashed lines in Fig.~\ref{fig:Smooth_Density}. The dividing lines have the same slope as the lines separating the BC/GV/RS regions, but with intercepts varying from $-6.5$ to $-9.5$ in steps $0.25$ wide. It is important to highlight that by adopting a slicing procedure rather than using three discrete zones, we largely remove the dependence of our results on the specific boundaries chosen for the BC/GV/RS. Nevertheless, despite being a somewhat arbitrary choice, the adopted boundary definition has proven to have physical meaning, since it is shown to coincide with the regions where most of the morphological transition happens \citep{2022MNRAS.509..567S}. Each zone corresponding to different vertical distances $\rm \Delta SFMS$ from the star-formation main sequence (SFMS; \citealt{2007ApJ...670..156D, 2007ApJ...660L..43N, 2011ApJ...730...61K}) defined by the line $\text{log(\rm SFR)} =  0.7 \log(M_\text{\rm stellar}) - 7.0$. The resulting variation in galaxy type fraction as a function of the vertical distance to the star-forming main sequence is shown in Fig.~\ref{fig:Dissection_SFR_Mstellar_Fraction} for different environments and stellar-mass ranges. As before, we limit our analysis to bins with at least 20 galaxies.
\subsection{Red but not dead: AGN in massive central red-sequence galaxies in groups}
\label{sec:AGN_in_ellipticals}

ETGs are mainly found in the red sequence and are usually assumed to be passively evolving, such that their star-formation has ceased and their stellar populations simple age with time. They are often called ``red and dead''. However, by tracing the Seyfert fraction as a function of $\rm \Delta SFMS$, we find that, depending on environment, not all central galaxies in the red sequence are quite as dead as previously thought. In panel (h) of Fig.~\ref{fig:Dissection_SFR_Mstellar_Fraction} we find that a relatively high fraction (about one third) of the most massive group galaxies with the lowest star-formation activity are AGN (panel h). This high AGN frequency is in agreement with our previous results, suggesting that massive central galaxies in groups (usually ellipticals) can be fed with gas via interactions with satellite galaxies, generating AGN activity.

\begin{figure}
    \centering
    \includegraphics[width = \columnwidth]{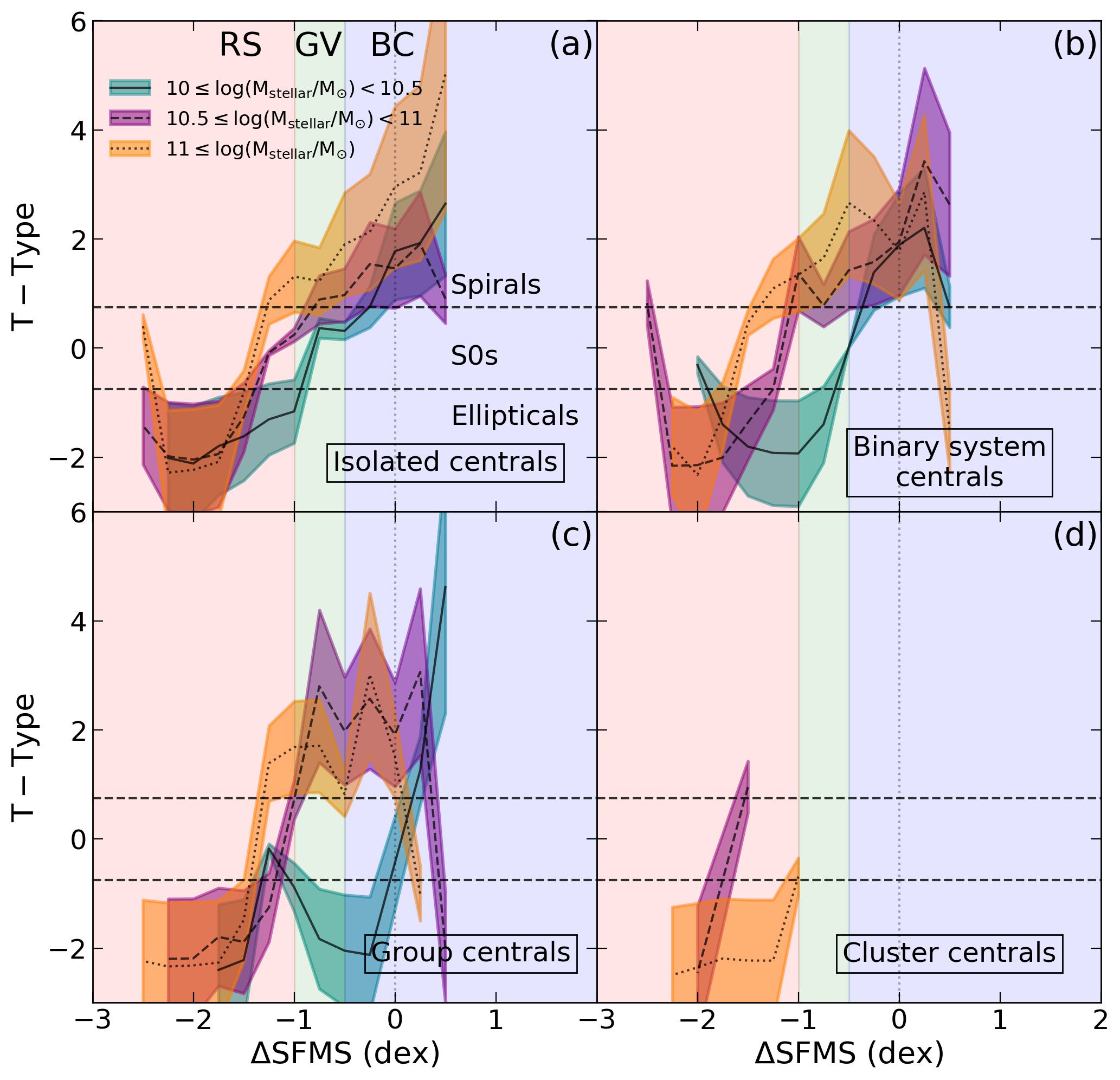}
   \caption{Median T-Type as a function of $\rm \Delta SFMS$ for central galaxies classified as Seyferts in the BPT diagram. We divide galaxies according to stellar mass (different colours and line styles) and environmental richness (different panels). Background colors correspond to the BC, GV and RS regions, as in Fig.~\ref{fig:Dissection_SFR_Mstellar_Fraction}. 1-sigma errors are shown as shaded areas.}
    \label{fig:Dissection_SFR_Mstellar_Morphology}
\end{figure}

\subsection{The AGN-driven transition from the blue cloud to the red sequence}
\label{sec:AGN_RS_BC}

The results shown in Fig.~\ref{fig:Dissection_SFR_Mstellar_Fraction} provide a new perspective on the possible effect of AGN activity on the evolution of central galaxies. The increasing Seyfert fraction across the BC as we move away from the SFMS coincides with a corresponding  decrease in the star-forming fraction, providing evidence that, despite representing only 10--20\% of the central galaxy population, AGN feedback could be extremely efficient at quenching star formation for all stellar masses. In this scenario, the AGN is responsible for decreasing the galaxies' star formation, but there is a time delay between the triggering of the AGN and its radiation becoming the main source of the ISM ionisation. During this process, the AGN activity becomes `unmasked' after the associated decrease in star formation reduces the ionisisng effect of the high-mass stars. High spatial resolution integral-field spectroscopic data from instruments such as MUSE may provide the necessary evidence to test this scenario.

These results are consistent with the idea that the transition from the BC to the RS is caused by AGN-driven outflows \citep{2020MNRAS.491.5406T}. Interestingly, the peak of the Seyfert fraction occurs in the vicinity of the GV, a region associated with both star-formation suppression and morphological transition \citep{2022MNRAS.509..567S}, suggesting, once again, that AGN activity is linked with galaxy transition from spirals to early-type galaxies.

The subsequent decrease in Seyfert fraction from the GV to the RS may be due to the combination of two processes, AGN feedback ejecting gas from central galaxies, which depletes the gas fuel reservoir needed to feed the SMBH, and the morphological transition from disks to spheroids, in which the gas is more stable and less likely to be accreted by the SMBH \citep{2009ApJ...707..250M}. In other words, the AGN itself is ultimately responsible for shutting down its own activity. 

\subsection{Morphological Transition vs. Star Formation Quenching}

By tracking the morphology of Seyfert galaxies as a function of their star-formation activity, we are also be able to address the question of whether AGN feedback changes the star formation activity of their host galaxies before or after their morphology changes. The median T-Type of Seyfert galaxies as a function $\rm \Delta SFMS$ is shown in Fig.~\ref{fig:Dissection_SFR_Mstellar_Morphology}, where the galaxies have been separated according to stellar mass and environmental richness. For group and cluster centrals, we show only the curves with a minimum number of 10 galaxies per bin to ensure the statistical uncertainties are not too large. 

This figure shows that, statistically, Seyfert galaxies reach elliptical morphologies (median T-Type $\rm \sim -2$) before their star-formation is completely suppressed (i.e., before they reach the core of the RS region, $\rm \Delta SFMS \sim -2$). There is also a mass dependence in the morphological transition of Seyfert galaxies: while low-mass galaxies typically emerge from the GV with elliptical morphologies (irrespective of their environmental richness), intermediate-mass systems emerge, on average, as S0s, and high mass systems tend to retain spiral morphologies beyond the GV. This provides evidence suggesting that the AGN activity is more relevant in the morphological transition of low-mass galaxies, whilst more massive systems hold their morphology ``longer''. 

In summary, our results show that morphological transition and star-formation suppression happen at different pace, in agreement with \cite{2022MNRAS.509..567S}. Although the detailed connection between AGN activity and the physical processes by which it affects morphological transition are still not wholly understood, it is noticeable that the sequencing we observe suggests that, irrespective of stellar mass and environment, the morphology of Seyfert galaxies evolves from late- to early-type before their star formation is fully quenched.

\section{Conclusions}

With the aim of shedding light understanding how AGN affect the evolution of their host galaxies, we have investigated the relationship between the properties of central galaxies, their environment, and the presence or absence of an AGN at their centres. We focus on central galaxies because, as the dominant object in their dark-matter halo, external influences may be weaker than for satellite galaxies, making them ideal candidates for studying the effect of internal processes such as AGN feedback.   

We have traced the effect of the environment by dividing the galaxies into isolated centrals and centrals in pairs, groups, and clusters. SDSS spectroscopy and the BPT diagram allow us to determine whether the dominant ionising mechanism of the galaxies' interstellar medium is star-formation or AGN activity. We analyse this information as a function of internal galaxy properties such as their stellar and dark-matter halo mass, morphology, and central velocity dispersion (correlated with their SMBH mass). In doing so we explore how the interplay between different combinations of internal and external factors may be linked to the fueling of the central SMBH and thus AGN activity, and how this activity may affect the properties and evolution of its host galaxies. The main results of this analysis are:

\begin{itemize}

    \item In all environments, the preeminence of AGN activity as the dominant ionising mechanism increases with stellar mass, overtaking star-formation for galaxies with $\rm M_\text{stellar} \geq 10^{11}M_\odot$ (Figs.~\ref{fig:seyfert_overall_fraction} and \ref{fig:Fract_vs_TType}). It is noticeable that, at these high masses, the fraction of LTGs with Seyfert activity reaches $\rm 21 \pm 4\%$ in group centrals, while only $\rm 7 \pm 3\%$ of these galaxies have their interstellar medium ionised by star formation.
    
    \item In high stellar mass central galaxies, the fraction of Seyferts increases with T-Type from $\rm 5 \pm 1\%$ at T-Type$\,\sim-2.5$ (ellipticals) to $\rm 18 \pm 3\%$ at T-Type$\,\sim4$ (late spirals). This increase in Seyfert fraction with morphology suggest that AGN feedback is linked with the morphology of these galaxies and, perhaps, its transformation. The low fraction of Seyfert nuclei in low-mass galaxies, independent of morphology and environment, results from the inability of these systems to fuel star-formation and AGN simultaneously.

    \item The separate variation of the Seyfert fraction with the galaxies' velocity dispersion and halo mass suggests two different ways to fuel AGN activity. First, central galaxies can rely on their own host halo gas reservoir to fuel their AGN, which is increasingly efficient with increasing halo mass; and second, central galaxies can increase their gas reservoir via interactions with satellite systems. We find that the second mechanism is most important for the evolution of high-mass centrals in groups and low-mass clusters, where such interactions will be more common. 

    \item The Seyfert fraction changes systematically as a function of star-formation activity in their host galaxies (Fig.~\ref{fig:Dissection_SFR_Mstellar_Fraction}). Within the blue cloud this fraction increases as star-formation activity declines, reaching a maximum when the galaxies enter the green valley. The peak fraction depends on both stellar mass and environmental richness. For instance, high mass group centrals show a peak Seyfert fraction of $\rm 18 \pm 4\%$ near the green valley. Subsequently, the Seyfert fraction decreases as the galaxies transition into the red sequence. This sequence strongly suggest that AGN feedback plays a key role in regulating and suppressing star-formation.     

    \item Tracing the morphology of Seyfert central galaxies as a function of their star-formation activity, we find that they typically achieve an early-type morphology  while they still host some residual star formation. This suggests that, in all environments, the morphology of Seyfert galaxies evolves from late- to early-type before their star formation is fully quenched. We additionally show that stellar mass plays an important role in the morphological evolution of Seyfert galaxies: low mass systems tend to emerge from the green valley with an elliptical morphology, whereas their high-mass counterparts maintain a spiral morphology deeper into the red sequence.

\end{itemize}

In summary, we have found strong evidence that AGN activity in central galaxies is intimately linked with the morphology and star-formation of their hosts, and that this link depends on both the internal properties of these galaxies and their environment. 

\section*{Acknowledgements}

VMS and RRdC acknowledge the support from FAPESP through the grants 2020/16243-3, 2020/15245-2 and 2021/13683-5. SZ, MRM, and AAS acknowledge financial support from the UK Science and Technology Facilities Council (STFC; grant ref: ST/T000171/1). IF acknowledges support from the Spanish Ministry of Science, Innovation and Universities (MCIU), through grant PID2019-104788GB-I00.

\section*{Data Availability}
 
The data underlying this paper were accessed from SDSS DR16 database
(https://skyserver.sdss.org/dr16/en/home). The data underlying this article will be shared on request to the corresponding author.



\bibliographystyle{mnras}
\bibliography{reference} 




\appendix

\section{Comparison using alternative diagnostic diagrams}
\begin{figure*}
    \centering
    \includegraphics[width = \textwidth]{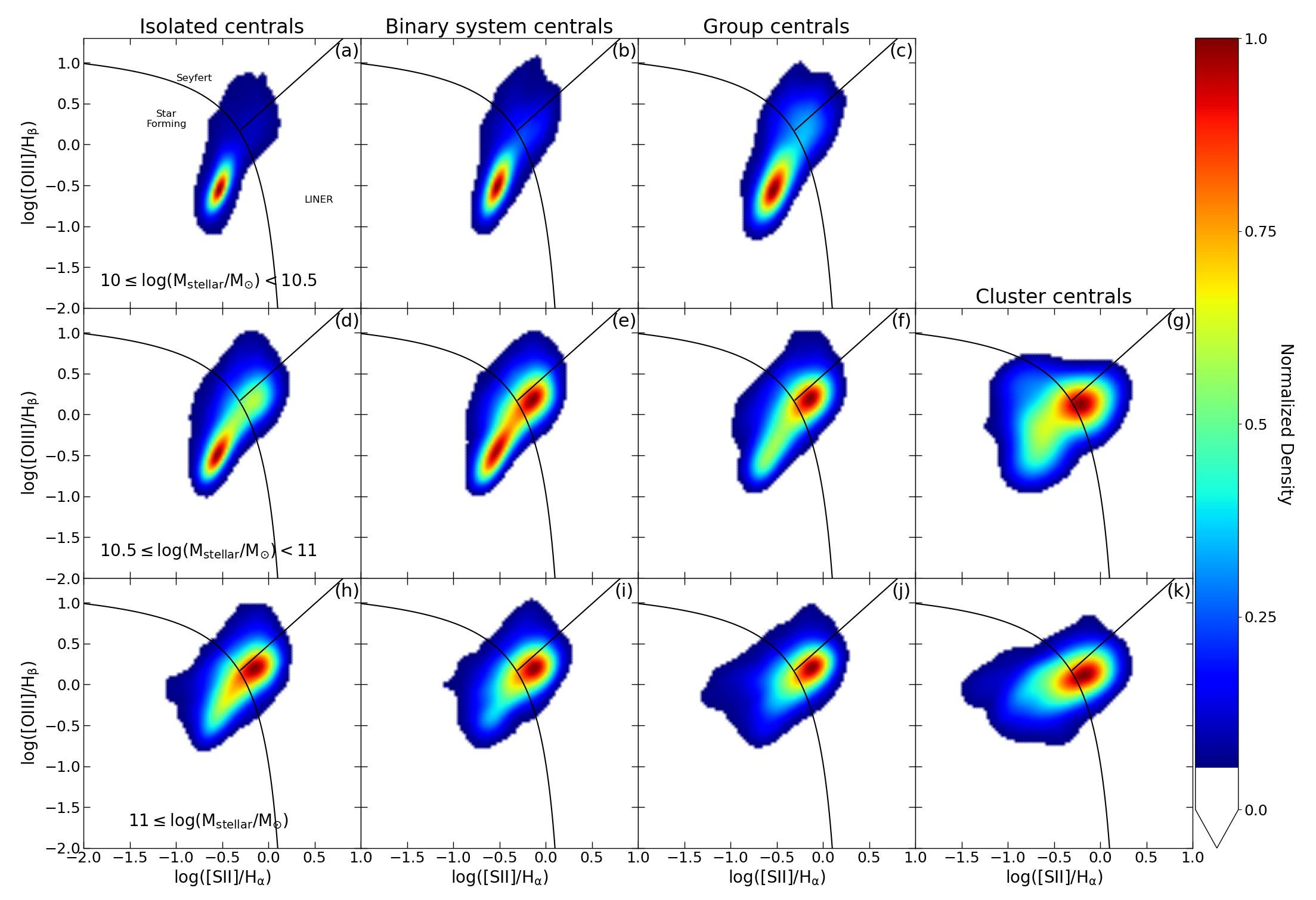}
    \caption{Gaussian kernel smoothed distribution of galaxies in the NII-BPT diagram. We divide systems according to stellar mass and environmental richness.}
    \label{fig:BPT_SII}
\end{figure*}

The main underlying hypothesis of this work is that the BPT diagram we have chosen is able to reliably identify the main mechanism ionizing the insterstellar gas in central galaxies. However, the literature provides alternative diagnostic diagrams to achieve the same goal using a variety of emission-line ratios. Since the question ``which diagnostic diagram is the best for this purpose?'' doesn't have a simple answer (see \citealt{2010MNRAS.403.1036C} for a comprehensive comparative study of several diagnostic diagrams), we have tested whether our findings depend on the specific diagnostic diagram we choose. As an example, we show in Fig.~\ref{fig:BPT_SII} the distribution of galaxies in an alternative BPT diagram that uses the [SII]/H$\alpha$ ratio instead of [NII]/H$\alpha$, as we did in Fig.~\ref{fig:BPT_Smooth}. Comparing both figures we find that the galaxy distribution is similar using the alternative diagnostic. Therefore, we are confident that our results do not depend significantly on the specific diagnostic diagram adopted.



\bsp	
\label{lastpage}
\end{document}